\documentclass[final,5p,times,twocolumn,number]{elsarticle}

\biboptions{square,sort&compress}

\usepackage{hyperref}
\hypersetup{
    colorlinks=true,
    linkcolor=red,
    filecolor=magenta,      
    urlcolor=blue,
    citecolor = teal,
}

\usepackage{amssymb}
\usepackage{lipsum}
\usepackage{amsmath}

\usepackage{color}
\usepackage{xcolor}
\usepackage{graphicx}

\definecolor{limegreen}{RGB}{154,205,0}
\definecolor{brickred}{rgb}{0.8, 0.25, 0.33}
\definecolor{amethyst}{rgb}{0.6, 0.4, 0.8}
\definecolor{azure}{rgb}{0.0, 0.5, 1.0}
\definecolor{awesome}{rgb}{1.0, 0.13, 0.32}
\usepackage[normalem]{ulem}
\newcommand{\replace}[2]{{{\color{violet}{#1}}{\color{limegreen}{\ifmmode\text{\sout{\ensuremath{#2}}}\else\sout{#2}\fi}}}}
\newcommand{\replacea}[2]{{{\color{purple}{#1}}{\color{olive}{\ifmmode\text{\sout{\ensuremath{#2}}}\else\sout{#2}\fi}}}} 
\newcommand{\replacem}[2]{{\color{azure}{#1}}{\color{brickred}{\ \ifmmode\text{\sout{\ensuremath{#2}}}\else\sout{#2}\fi}}}

\newcommand{\avr}[1]{\left\langle #1 \right\rangle}
\newcommand{\aavr}[1]{\left\llangle #1 \right\rrangle}
\newcommand{\eq}[1]{(\ref{#1})}
\newcommand{\bs}{\boldsymbol}

\newcommand{\Tr}{\mbox{ Tr}}

\makeatletter
\DeclareFontFamily{OMX}{MnSymbolE}{}
\DeclareSymbolFont{MnLargeSymbols}{OMX}{MnSymbolE}{m}{n}
\SetSymbolFont{MnLargeSymbols}{bold}{OMX}{MnSymbolE}{b}{n}
\DeclareFontShape{OMX}{MnSymbolE}{m}{n}{
    <-6>  MnSymbolE5
   <6-7>  MnSymbolE6
   <7-8>  MnSymbolE7
   <8-9>  MnSymbolE8
   <9-10> MnSymbolE9
  <10-12> MnSymbolE10
  <12->   MnSymbolE12
}{}
\DeclareFontShape{OMX}{MnSymbolE}{b}{n}{
    <-6>  MnSymbolE-Bold5
   <6-7>  MnSymbolE-Bold6
   <7-8>  MnSymbolE-Bold7
   <8-9>  MnSymbolE-Bold8
   <9-10> MnSymbolE-Bold9
  <10-12> MnSymbolE-Bold10
  <12->   MnSymbolE-Bold12
}{}

\let\llangle\@undefined
\let\rrangle\@undefined
\DeclareMathDelimiter{\llangle}{\mathopen}%
                     {MnLargeSymbols}{'164}{MnLargeSymbols}{'164}
\DeclareMathDelimiter{\rrangle}{\mathclose}%
                     {MnLargeSymbols}{'171}{MnLargeSymbols}{'171}
\makeatother

\allowdisplaybreaks

\journal{Physics Letters B}

\begin{document}

\begin{frontmatter}



\title{Negative moment of inertia and rotational instability of gluon plasma}


\author[bltp,mipt]{Victor V. Braguta}
\ead{vvbraguta@theor.jinr.ru}
\author[tours]{Maxim N. Chernodub}
\ead{maxim.chernodub@univ-tours.fr}
\author[bltp,dubna]{Artem A. Roenko}
\ead{roenko@theor.jinr.ru}
\author[bltp,mipt]{Dmitrii A. Sychev}
\ead{sychev.da@phystech.edu}
\affiliation[bltp]{organization={Bogoliubov Laboratory of Theoretical Physics, Joint Institute for Nuclear Research},
            city={Dubna},
            postcode={141980}, 
            country={Russia}}
\affiliation[mipt]{organization={Moscow Institute of Physics and Technology},
            city={Dolgoprudny},
            postcode={141700}, 
            country={Russia}}
\affiliation[tours]{organization={Institut Denis Poisson UMR 7013, Université de Tours},
            city={Tours},
            postcode={37200}, 
            country={France}}
\affiliation[dubna]{organization={Dubna State University},
            city={Dubna},
            postcode={141980}, 
            country={Russia}}

\begin{abstract}
Using first-principle numerical simulations of the lattice SU(3) gauge theory, we calculate the isothermal moment of inertia of the rigidly rotating gluon plasma. We find that the moment of inertia unexpectedly takes a negative value below the ``supervortical temperature''  $T_s = 1.50(10) T_c$, vanishes at $T = T_s$, and becomes a positive quantity at higher temperatures. The negative moment of inertia indicates a thermodynamic instability of rigid rotation. We derive the condition of thermodynamic stability of the vortical plasma and show how it relates to the scale anomaly and the magnetic gluon condensate. The rotational instability of gluon plasma shares a striking similarity with the rotational instabilities of spinning Kerr and Myers-Perry black holes.  
\end{abstract}




\end{frontmatter}




%
%
\section{Introduction}
\label{sec:intro}

In thermodynamic equilibrium, all physical objects have positive moments of inertia, implying that in order to set a static object into rotation, one needs to apply a torque~\cite{LL1}. While a negative moment of inertia is impossible in thermal equilibrium, this effect can be achieved in non-equilibrium, open systems. 

In mechanics, the realization of a $I<0$ system requires the presence of an active component such as a motor~\cite{Lonar2022}. In electronics, the relevant example is played by electrical negative-impedance converters identified as an active electric circuit with negative resistivity~\cite{Chen2002-wo}. A negative moment of inertia can also be realized in rotating Casimir systems associated with negative vacuum energy~\cite{Chernodub:2012ry, Chernodub:2012em, Flachi:2022hfj}. In addition, the negativity of isothermal moment of inertia can be achieved in thermodynamically unstable systems such as rotating black holes~\cite{Whiting1988, Prestidge2000, Reall2001, Monteiro_2009tc, Altamirano2014}. 

In our paper, we show that the {\it rigidly} rotating gluon plasma possesses, in thermal equilibrium, a negative moment of inertia ($I < 0$) below the temperature
\begin{align}
    T_s = 1.50\,(10) \, T_c\,,
\label{eq_T_s}
\end{align}
where $T_c$ is the deconfining transition temperature in the non-rotating plasma. We call $T_s$ the ``supervortical temperature'' since at $T = T_s$, the {\it rigidly} rotating plasma loses its moment of inertia, $I(T_s) = 0$, in a distant similarity with a superconductor which loses its resistivity at a certain critical temperature.

Rotating quark-gluon plasma (QGP) with temperatures around the supervortical temperature~\eq{eq_T_s} is routinely produced in relativistic heavy-ion collisions. Such plasma can have exceptionally high vorticity of the order of $\omega$~$\approx (9 \pm 1)\times 10^{21} \, \mathrm{s}^{-1} \sim 0.03\, \mathrm{fm}^{-1} c \sim 7 \, \mathrm{MeV}$~\cite{STAR_2017ckg}. The properties of vortical QGP can be probed via spin polarization of produced hadrons that provide us with an opportunity to confront theoretical methods with experimental results~\cite{Becattini_2020ngo, Huang:2020dtn}. While the rotation in the expanding plasma fireball is not a solid rotation, all theoretical -- both numerical and analytical -- approaches to the thermodynamics of rotating QGP assume a rigid rotation of the system, which drastically simplifies analytical treatment of the problem~\cite{Huang:2020dtn, Ambrus_2014uqa, Ambrus_2015lfr, Chen_2015hfc, Jiang_2016wvv, Chernodub_2016kxh, Chernodub_2017ref, Wang_2018sur, Zhang_2020hha, Sadooghi_2021upd, Chen_2020ath, Fujimoto_2021xix, Golubtsova:2021agl, Chen:2022smf, Golubtsova:2022ldm, Zhao:2022uxc, Chernodub_2020qah, Chernodub_2022veq, Braguta_2020biu, Braguta_2021jgn, Braguta_2022str}, allowing, the same time, to probe the effect of vorticity on thermodynamics of the system in a systematic and controllable way and to tackle the approach to the thermodynamic limit (see, in particular, Refs.~\cite{Chernodub_2017ref, Braguta_2021jgn, Chen:2022smf}).

The thermal transition from hadronic to the QGP phase is accompanied by the restoration of the chiral symmetry and the deconfinement of color. There is a general agreement in the community that the rigid rotation, according to all model estimates, should reduce the critical temperature of the chiral transition in the fermionic sector~\cite{Chen_2015hfc, Jiang_2016wvv, Chernodub_2016kxh, Chernodub_2017ref, Wang_2018sur, Zhang_2020hha, Sadooghi_2021upd}. 

However, the situation with the deconfining transition is not clear: the rigid rotation should either drive plasma to the deconfinement phase~\cite{Chen_2020ath, Fujimoto_2021xix, Golubtsova:2021agl, Chen:2022smf, Golubtsova:2022ldm, Zhao:2022uxc} or, with another scenario, should not affect the system at the rotational axis, forming, at high vorticity, an inhomogeneous confining-deconfining phase (the inverse hadronization effect)~\cite{Chernodub_2020qah}. While signatures of the inhomogeneity are seen in kinematic variables in numerical simulations of pure gluon plasma~\cite{Chernodub_2022veq}, the numerical first-principle simulations have also revealed that the bulk critical temperature of the deconfining phase transition grows with the increase of the angular frequency~\cite{Braguta_2020biu, Braguta_2021jgn}. Moreover, it turns out that gluons and fermions have opposite effects on the critical temperature in rotating QGP. It seems that the gluon sector wins in this contest, and the deconfinement, as well as the chiral critical temperatures, increase with the rotation~\cite{Braguta_2022str}.
The results for the behavior of critical temperature in rotating QCD~\cite{Braguta_2022str} have also been recently confirmed using another fermionic lattice action~\cite{Yang:2023vsw}.

Thus, the model-based analytical approaches and the first-principle numerical simulations of rigidly rotating gluon plasma do not match. To explore this puzzle deeper, in our work we look at the mechanical properties of the rotating gluon plasma.

%
%
\section{Angular momentum and moment of inertia}

A mechanical response of a thermodynamic ensemble to a rigid rotation with the angular velocity $\bf \Omega$ can be quantified in terms of the conjugated variable, the total angular momentum $\bs J$, which includes orbital and spin parts. These quantities determine the relation between the energy $E^{\text{(lab)}}$ in the inertial laboratory frame and the energy in the co-rotating, non-inertial reference frame, $E = E^{\text{(lab)}} - {\bs J} {\bs \Omega}$~\cite{LL1}.
The angular momentum ${\bs J}$ can be expressed via either the energy $E$ or the free energy $F = E - T S$ in the co-rotating frame:
\begin{align}
{\bs J} = - {\left( \frac{\partial E}{\partial {\bs \Omega}} \right)}_{S} = - {\left( \frac{\partial F}{\partial {\bs \Omega}} \right)}_{T}\,,
\label{eq_L_via_tildeF}
\end{align}
where we used $d E = T d S - {\bs J} d {\bs \Omega}$ and 
    $d F = - S d T - {\bs J} d {\bs \Omega}$. 
    
The moment of inertia is a scalar quantity,
\begin{align}
I(T,\Omega) = \frac{J(T,\Omega)}{\Omega} = - \frac{1}{\Omega} {\left( \frac{\partial F}{\partial \Omega} \right)}_{T}\,,
\label{eq_moment_inertia}
\end{align}
which fixes a relation between the angular momentum ${\bs J}(T,\Omega) = I(T,\Omega) {\bs \Omega}$ and the angular velocity ${\bs \Omega} = \Omega {\bf e}$ of rotation around a fixed axis ${\bf e}$.

We start our discussion with a cylinder-shaped gluon plasma with a radius~$R$, rigidly rotating with the angular frequency $\Omega$ around the symmetry axis. We consider slowly rotating gluon plasma implying the velocity,
\begin{align}
    v_R = \Omega R\,,
    \label{eq_V_R}
\end{align}
at the boundary is non-relativistic, $v_R^2 \ll 1$, thus ensuring that the rotating system respects the causality bound, $-1 < v_R < 1$. It is important to notice that for rotating quantum fields, two possible extreme types of vacua have been considered in the literature: the nonrotating (Vilenkin) vacuum~\cite{Vilenkin:1980zv} and the rotating (Iyer) vacuum~\cite{Iyer:1982ah}. For a causally rotating vacuum, these vacua are identical and, therefore, a thermodynamic system built over these vacua has the same physical properties ({\it e.g.}, phase diagram or thermodynamic instabilities) in all frames~\cite{Ambrus_2015lfr}.

As we work with a large system size, $R \sim \mbox{(a few) fm}$, the condition of the slowness of rotation implies also that the angular velocity is much smaller than the intrinsic QCD energy scale, $\Omega \ll \Lambda_{\mathrm{QCD}}$.
Moreover, in the whole range of temperatures in our work, $T \simeq (1.0 \sim 2.0) T_c$, the boundary effects can also be neglected because the spatial thermal correlation lengths in the strongly interacting gluonic plasma at $T \gtrsim T_c$ are of the order of $\Lambda_{\mathrm{QCD}}^{-1}$ or shorter. In contrast, below $T_c$, the correlations are governed by the glueball masses, which correspond to even shorter correlation lengths, e.g. $M_{0^{++}} = 1.653(26) \, \mathrm{GeV}$~\cite{Athenodorou_2020ani}. These physical conditions (system size $R$, temperature range $T$, and rotational frequency range $\Omega$) correspond to physical conditions of vortical plasma created at RHIC in noncentral relativistic heavy-ion collisions~\cite{STAR_2017ckg}.

%
%
\section{Thermodynamics and velocity at the boundary.}\label{sec_3}

Several theoretical approaches to rigidly rotating gluon plasma~\cite{Huang:2020dtn, Fujimoto_2021xix} express its thermodynamic properties as a function of the angular frequency $\Omega$, suggesting independence (or a mild dependence) of the thermodynamics on the system size in directions perpendicular to the axis of rotation~\cite{Fujimoto_2021xix}. However, first-principle numerical simulations indicate that the transverse size dependence is very pronounced~\cite{Braguta_2020biu, Braguta_2021jgn} (see also discussion in Ref.~\cite{Chernodub_2016kxh} and the explicit derivation for a bosonic system in Ref.~\cite{Ambrus:2023bid}). Moreover, it appears that the thermodynamic potentials of the slowly rotating system incorporate angular frequency only via the common product $\Omega R$ given by Eq.~\eq{eq_V_R}.\footnote{I.e., pressure does not depend on the combination $\Omega/ \Lambda_{\mathrm{QCD}}$.}

We neglect a shape change of the slowly rotating plasma, thus assuming that the mass density $\rho_0$, which represents the number of degrees of freedom that couple to rigid rotation, is a coordinate- and $\Omega$-independent quantity~\cite{LL5}. Then the moment of inertia does not depend on $\Omega$:
\begin{align}
    I(T,R) 
    = \frac{\pi}{2} L_z R^4 \rho_0(T)
    \equiv - K_2(T) F_0(T,R) R^2\,,
\label{eq_I_T_Omega}
\end{align}
where $F_0 \equiv F^{\text{(lab)}}(\Omega = 0) < 0$ is the free energy of the non-rotating gas. The dimensionless coefficient $K_2$ in Eq.~\eq{eq_I_T_Omega} has a sense of a specific moment of inertia. It is a non-extensive quantity that corresponds to the moment of inertia of the system normalized by the system size and its free energy in the non-rotating state. The expected insensitivity of this quantity on the size of the system is particularly useful as it allows us to check the volume independence of our results.

The thermodynamic relation~\eq{eq_moment_inertia} implies that the co-rotating free energy,
\begin{subequations}
    \begin{align}
        F(T, R, \Omega) & = F_0(T, R) - \frac{1}{2} I(T,R) \Omega^2 
        \label{eq_F_series} \\
        & \equiv F_0(T,R) \Bigl(1 + \frac{1}{2} K_2(T) v_R^2 
        \Bigr)\,,
        \label{eq_F_vR}
    \end{align}
\label{eq_F_all}
\end{subequations}
possess a minus sign for the $\Omega^2$ term as it represents a centrifugal energy responsible for particle run-away forces directed outwards the axis of rotation.

%
%
\section{Numerical first-principle results}
We calculate the free energy density $f = F/V$ using the standard relations~\cite{Boyd_1996bx} 
\begin{subequations}
\label{eq_F_S_lat} \label{eq_free_energy}
\begin{align}
    \frac{f(T)}{T^4} & = - N_t^4 \int_{\beta_0}^{\beta} d \beta' \Delta s(\beta')\,,
    \label{eq_F_lattice} \\
    \Delta s(\beta) & = \langle s(\beta) \rangle_{T=0} - \langle s(\beta) \rangle_{T},
    \label{eq_Delta_S}
\end{align} 
\end{subequations}
where the $\mathrm{SU}(N_c)$ lattice coupling $\beta = 2 N_c/g_0^2$
is expressed in terms of the bare continuum coupling $g_0$, while $s=T/V\cdot S$ is the density of the lattice action $S$.

For Euclidean lattice calculations, the angular velocity $\Omega$
is put in the purely imaginary form $\Omega_I = - i \Omega$ to avoid the sign problem~\cite{Yamamoto_2013zwa}. The expressions for thermodynamic equilibrium in the Minkowski spacetime can be obtained by the analytic continuation. In particular, the velocity $v_R$ at the boundary~\eq{eq_V_R} becomes imaginary $v_I = - i v_R$, with the following relation:
\begin{equation}
    v_I^2 = - v_R^2\,.
\label{eq_V_I_R}
\end{equation}

%
%
\begin{figure}[t]
\centering
  \includegraphics[width=0.99\linewidth]{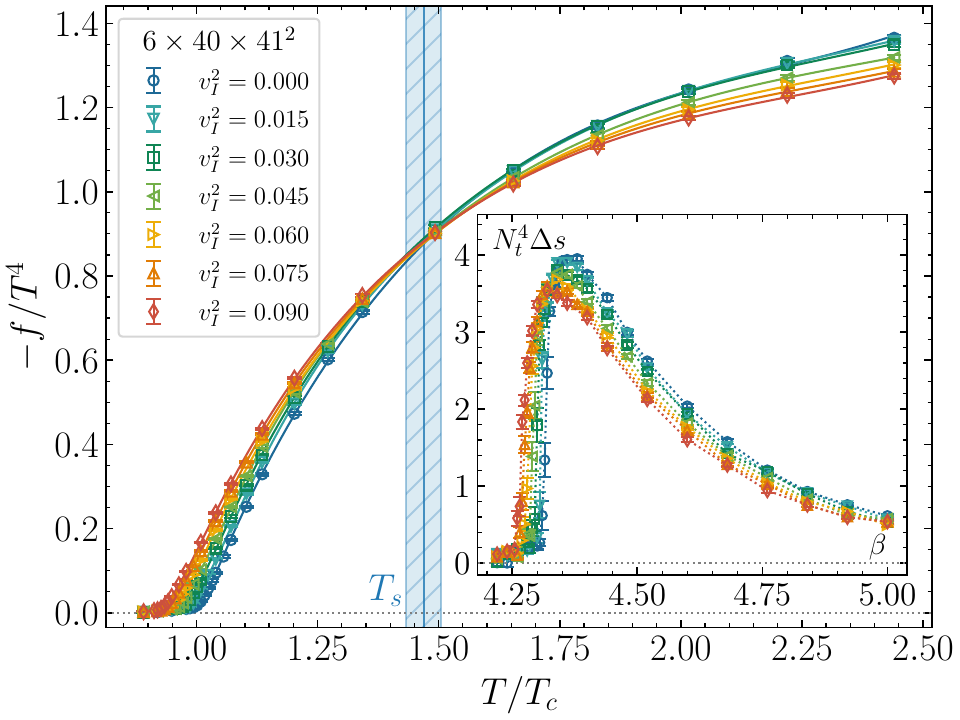}
  \caption{The free energy density $f$ in the co-rotating frame as a function of the temperature $T$ for the {$N_t = 6$} lattice. The vertical line shows the supervortical temperature $T_s$ for this lattice. The inset shows the expectation value of the lattice action density $\Delta s$ 
  as a function of the lattice gauge coupling~$\beta$. Both plots are given for several imaginary velocities squared $v_I^2$ at the boundary 
  of the system.
  }
  \label{fig_Delta_S}
\end{figure}

Our calculations are performed on the lattices of size $N_t \times N_z \times N_s^2 = N_t \times 40 \times 41^2 $ with $N_t = 5, 6, 7, 8$
and $N_t = 40$ for the zero temperature subtraction in Eq.~\eqref{eq_Delta_S}.
We incorporate the rotation into the lattice simulations following Ref.~\cite{Yamamoto_2013zwa, Braguta_2021jgn} using the tree-level improved Symanzik gauge action~\cite{Curci:1983an, Luscher:1985zq}. Other details about our lattice setup are provided in the Supplementary Material. We use periodic boundary conditions in all directions\footnote{Our previous study shows that lattice simulations of rotating gluodynamics with different boundary conditions give qualitatively the same results~\cite{Braguta_2021jgn}.}.
The imaginary velocity at the boundary is identified with the velocity at the middle of the boundary side, $v_I = \Omega_I R$, where ${R = a(N_s - 1)/2}$
is the distance from the boundary to the rotational axis, which is the $z$-axis.
In our work, we keep the angular velocity in lattice units constant as the value of lattice coupling $\beta$ varies. This condition ensures that the physical linear velocity $v_I$ of the plasma at the boundary of the system remains fixed during the integration in Eq.~\eq{eq_F_S_lat} as temperature varies.

\begin{figure}[t]
\centering
  \includegraphics[width=.99\linewidth]{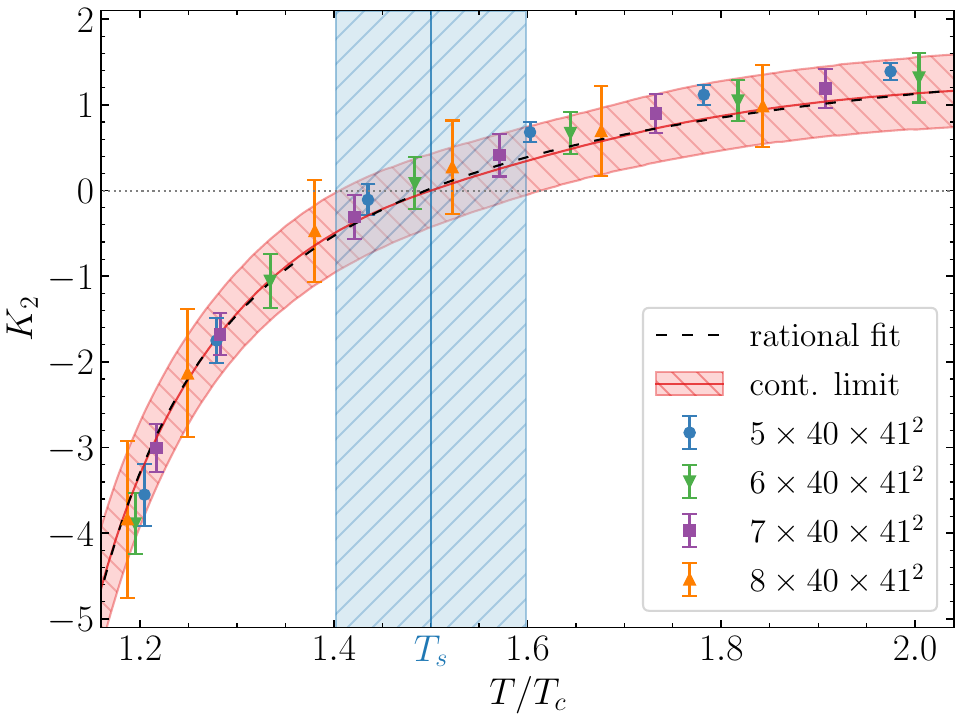}
  \caption{The dimensionless moment of inertia $K_2$ of the gluon plasma as a function of the temperature $T$ for {the used lattices}. 
  The red shaded region, with the central values marked by the red solid line, denotes the continuum extrapolation, $K_2 = K_2^{(\text{cont.})} + C/N_t^2$ at $N_t \to \infty$. 
  The best fit~\eq{eq_fit} of the continuum curve is shown by the dashed black line. The position of the supervortical temperature $T_s$ in the continuum limit is marked by the vertical line, which separates the unstable ($T < T_s$) and stable ($T > T_s$) regimes of rigid plasma rotation. The error estimations (the shaded regions for $T_s$ and $K_2^{(\text{cont.})}$, and the bars of the data) include both statistical and systematic uncertainties.
  }
  \label{fig_K2}
\end{figure}

Our numerical approach, which involves the integration of the lattice coupling~$\beta$ in Eq.~\eq{eq_F_S_lat}, respects the line of constant physics. This property follows from the fact, discussed in Sec.~\ref{sec_3}, that the angular frequency $\Omega$ enters the co-rotating free energy only in the combination $v_I=\Omega_I R$. Therefore, despite a variation of the lattice coupling $\beta$ affecting both the physical angular velocity $\Omega_I$ and the physical size of the system $R$, the system remains physically the same provided the product of these two quantities is kept constant.\footnote{Here, we silently require that the temperature is fixed and the system is sufficiently large so that the finite volume effects are small. The fulfillment of the former property in our simulations of the gluon plasma is seen from the excellent scaling of Fig.~\ref{fig_K2}, where points with different spatial sizes collapse to the same function of temperature, while the latter property has been thoroughly verified in our related study~\cite{Braguta:2023kwl}.}
While we cannot guarantee that the boundary velocity is not subjected to renormalization in the course of our procedure, the coincidence of our results with the ones obtained on a non-rotating lattice with a completely different method --shown in Fig.~\ref{fig_K2comp} and discussed in detail below-- strongly suggests that the renormalization effects are within the uncertainty of the calculation.

In the inset of Fig.~\ref{fig_Delta_S}, we show the normalized difference of lattice action densities~$\Delta s$, Eq.~\eq{eq_Delta_S}, which enters the free energy density~\eq{eq_F_lattice}. At vanishing velocity of the rotation, $v_I = 0$, we recover the known result~\cite{Boyd_1996bx, Borsanyi:2012ve, Beinlich:1997ia}. The steep rise of $\Delta s$, which happens close to the critical coupling $\beta \simeq \beta_c$, points to the first-order nature of the phase transition in the non-rotating plasma.

As the imaginary velocity $v_I$ increases, the transition shifts towards smaller lattice couplings $\beta$, signaling that the critical temperature $T_c = T_c(v_I)$ decreases as the {\it imaginary} angular frequency $\Omega_I$ (the velocity $v_I$ of the rotation) raises. This result is in agreement with previous numerical calculations~\cite{Braguta_2020biu, Braguta_2021jgn, Braguta_2022str}. 

The normalized free energy density in the co-rotating frame, $ - f/T^4$, calculated via Eq.~\eq{eq_F_lattice}, is shown in Fig.~\ref{fig_Delta_S}. This quantity is a monotonically raising function of the temperature $T$ at all imaginary velocities $v_I$, indicating the presence of a plateau at $T \to \infty$ for each fixed~$v_I$. 

%
%

The free energy density $f$, shown in Fig.~\ref{fig_Delta_S}, is a rising (diminishing) function of $v_I^2$ at fixed temperature $T < T_s$ ($T > T_s$). This property can be quantified by fitting the free energy density with a parabolic function of $v_I$:
\begin{align}
    f(T,v_I)  = f_0(T) \Bigl(1 - \frac{1}{2} K_2(T) v_I^2 \Bigr)\,,
\label{eq_F_v_I}
\end{align}
where $f_0$ and $K_2$ serve as fit parameters (with $f_0 < 0$). The expression in the Euclidean spacetime~\eq{eq_F_v_I} corresponds to the free energy~\eq{eq_F_vR} in the co-rotating frame in the Minkowski spacetime after the Wick transformation for the boundary velocities~\eq{eq_V_I_R}. The dimensionless moment of inertia $K_2$ is shown in Fig.~\ref{fig_K2} for the used lattices and in the continuum limit ($a\to 0$, or, equivalently, $1/N_t\to 0$ at a fixed temperature $T$). 

A striking feature of the free energy, Fig.~\ref{fig_Delta_S}, is that the curves corresponding to different $v_I$ intersect at the same ``supervortical'' temperature $T_s$, signaling that at this temperature, the free energy~\eq{eq_F_v_I} loses, at least for slow rotation $v_I^2 \ll 1$, the dependence on the rotational frequency. Therefore, the rigidly rotating gluon plasma loses its moment of inertia at $T = T_s$. We use the $K_2(T_s) = 0$ property as a definition of the supervortical temperature, and show its continuum limit in Fig.~\ref{fig_K2}. 

The continuum limit of the dimensionless moment of inertia $K_2$ can be well reproduced by a rational function 
\begin{align}
    K_2^{\mathrm{(fit)}}(T) = K_2^{(\infty)} - \frac{c}{T/T_c -1}\,, 
\label{eq_fit}
\end{align}
where the best-fit parameters are the high-temperature asymptotics $K_2^{(\infty)} = 2.23(39)$ and the slope coefficient $c = 1.11(20)$. The high-$T$ limit of the moment of inertia, $K_2^{(\infty)}$, is a non-universal quantity that may depend on the geometry of the system. 

%
%
\section{(In)stability and sign of the moment of inertia}

Formally, the negative moment of inertia, observed in the region of temperatures $T < T_s$, might imply that the rotation causes the quark-gluon plasma to cool down (the faster the rotation, the lower the thermal energy). However, physically, this counter-intuitive effect, contradicting the kinematic Tolman-Ehrenfest picture~\cite{Tolman_1930ona, Tolman_1930zza}, indicates that the rigid rotation is impossible thermodynamically, thus making the physical rotation non-rigid. Similar instabilities occur in curved gravitational backgrounds of rotating Kerr and Myers-Perry black holes~\cite{Whiting1988, Prestidge2000, Reall2001}.

For a system in stable equilibrium at a given temperature $T$ and angular velocity $\bs \Omega$, any deviation from the equilibrium should obey the following condition~\cite{LL5}:
\begin{align}
    \delta E - T \delta S - {\bs \Omega} \delta {\bs J} > 0\,,
\end{align}
which implies that all eigenvalues of the inverse Weinhold metric, defined in the thermodynamic space~\cite{Weinhold1975},
\begin{align}
    g^{(W),\mu\nu} = - \frac{\partial^2 f(T,{\bs \Omega})}{\partial X_\mu \partial X_\nu}\,, \qquad X_\mu = (T, \Omega_i)\,,
    \label{eq_gW}
\end{align}
must be positively defined (see a discussion in~\cite{Monteiro_2009tc}).

The positivity of the matrix~\eq{eq_gW} is achieved provided the specific heat at constant angular momenta $C_J = T \left(\partial S/\partial T \right)_J$ and the eigenvalues (spectrum) of the tensor of isothermal differential moment of inertia $I^{ij}  \equiv I^{ji} = \left( \partial J^i/\partial \Omega_j \right)_{T}$ are positive quantities: $C_J > 0$ and ${\mathrm{spec}}(I^{ij}) > 0$, respectively. The former is a standard requirement for the thermodynamic stability~\cite{LL5}, while the latter, given the (square) cylindrical geometry of our system, is reduced to the requirement $I > 0$ for the principal moment of inertia~\eq{eq_moment_inertia} at infinitesimally slow rotations, $\Omega \to 0$. In terms of the coefficient $K_2$, Eq.~\eq{eq_F_vR}, the thermodynamic stability thus requires:
\begin{align}
    K_2(T) > 0 \qquad\ \text{(thermodynamic stability)},
\end{align}
which is violated below the supervortical temperature, $T<T_s$. This instability has a thermodynamic origin. It has no obvious relation to hydrodynamic instabilities that might be generated by the viscous flow of hot gluons. Moreover, theoretically, instability is expected to be realized only at near-luminal velocities, $v_R \to 1$~\cite{Braga:2023qee}.

%
%
\section{Moment of inertia and scale anomaly}

The difference of lattice action densities $\Delta s$, shown in the inset of Fig.~\ref{fig_Delta_S}, is closely connected with the scale anomaly~\cite{Boyd_1996bx}. 
The moment of inertia $I$ is also directly related to the trace anomaly (see Supplementary Material):
\begin{equation}
    I(T) = - V T^4 \int_{0}^T \frac{d T'}{T'} \frac{\langle T^{\mu}_{\ \mu}\rangle^{(2)}(T')}{{T'}^4}\,, 
    \label{eq_I_int}
\end{equation}
where $\langle T^{\mu}_{\ \mu}\rangle^{(2)}$ is the second moment of the anomalous trace
\vspace{-.1em}
\begin{equation}
\langle T^{\mu}_{\ \mu}\rangle^{(2)}(T) = \left[ \frac{\partial^{2}}{\partial \Omega_I^{2}} \langle T^{\mu}_{\ \mu}\rangle(T,\Omega_I) \right]{\Biggl|}_{\Omega_I = 0}\,.
\label{eq_T_moments}
\end{equation}
Therefore, the moment of inertia corresponds to the rotational response of the scale anomaly.

%
%
\section{ Role of the magnetic gluon condensate}

Using Eqs.~\eqref{eq_moment_inertia} and \eqref{eq_F_series}, we get the gluon moment of inertia:
\begin{align}
    I & = I_{\text{mech}} + I_{\text{magn}} 
    = \frac{1}{T} \int_V d^3 x \int_V d^3 x' \left\llangle M^{12}_0({\bs x}) M^{12}_0({\bs x}')\right\rrangle_T \nonumber \\ 
   & {} + \int_V d^3 x \left\llangle(\epsilon^{ij} F^a_{i3} x_j)^2 + (F^a_{12})^2 (x^2_1 + x^2_2) \right\rrangle_T\,,  \label{eq_I_F}
\end{align}
where the relation $\Omega^2 = - \Omega^2_I$ is used. Here
\begin{align}
    M^{ij}_{0}({\bs x}) = x^i T^{j0}({\bs x}) - x^j T^{i0}({\bs x})\,, \qquad i,j = 1,2,3\,,
    \label{gl_ang_mom}
\end{align}
is the local angular momentum related to the $\Omega \to 0$ limit of the gluon stress-energy tensor:
$T^{\mu\nu} = F^{a,\mu\alpha} F^{a,\nu}_{\quad\, \alpha} - (1/4) \eta^{\mu\nu} F^{a,\alpha\beta} F^a_{\alpha\beta}$. 
We also denoted $\llangle {\mathcal O} \rrangle_T = \left\langle {\mathcal O} \right\rangle_T- \left\langle {\mathcal O} \right\rangle_{T=0}$ to represent the thermal part of the expectation value of an operator $\mathcal O$. The normalization of Eq.~\eq{eq_I_F} is chosen by requiring that the cold ($T = 0$) vacuum has no inertia.

The first term in Eq.~\eq{eq_I_F} corresponds to a mechanical term that describes fluctuations of the angular momentum via a standard linear response form
\begin{align}
    I_{\text{mech}} = 
    \frac{1}{T} \left\llangle (J^3)^2 \right\rrangle_T, \quad
    J^i = \frac{1}{2} \int_V d^3 x \, \epsilon^{ijk} M^{jk}_{0}({\bs x})\,,
\label{eq_I_corr}
\end{align}
where we used 
that $\big\langle{J^3}\big\rangle = 0$ at $\Omega = 0$ at any temperature.

The second term in~\eq{eq_I_F} involves a nonperturbative magnetic gluon condensate in the static, $\Omega \to 0$, limit. Using the $\mathrm{SO}(3)$ rotational symmetry and the translational invariance of the plasma in spatial dimensions, we get the relation $\big\llangle{F^a_{i3} F^a_{j3}}\big\rrangle_T = \delta_{ij} \big\llangle{(F^a_{12})^2}\big\rrangle_T$, which can be expressed via the magnetic gluon condensate at a finite temperature $\big\llangle{(F^a_{ij})^2}\big\rrangle_T \equiv 6 \big\llangle{(F^a_{12})^2}\big\rrangle_T$. We get:
\begin{align} \hskip -1mm
    I_{\text{magn}} = 
    \frac{1}{3}\int_V d^3 x \, x_\perp^2 \aavr{(F_{ij}^{a})^2}_{T} = \frac{\pi}{6} L_z R^4 \aavr{(F_{ij}^a)^2}_T\,.
\label{eq_I_cond}
\end{align}
Surprisingly, this relation has the same form as the classical formula for the moment of inertia~\eq{eq_I_T_Omega}, where the mass density $\rho_0$ corresponds to the thermal part of the magnetic gluon condensate $\rho_0 (T) = \big\llangle{(F_{ij}^a)^2}\big\rrangle_T/3$. 

The full gluon condensate $\avr{G^2}$ (a sum of its magnetic and electric parts) is a phenomenologically important quantity which takes a positive value at $T = 0$~\cite{Shifman_1978bx, Shifman_1978by}. It decreases monotonically with the increase of the temperature, implying that the thermal part of the condensate, $\aavr{G^2}_T$, always takes a negative value~\cite{Miller_1997dn, Gubler_2018ctz}. This ``melting'' of the gluon condensate agrees with the negative value of the thermal part of the scale anomaly: $\aavr{G^2}_T = - \avr{T^\mu_{\ \mu}}_T < 0$ (see a discussion in~\cite{Boyd_1996bx}). 

However, the magnetic contribution to the scale anomaly 
reverses its sign at $T \simeq 2 T_c$~\cite{Boyd_1996bx} indicating that the magnetic part of the thermal gluon condensate becomes positive and implying that $I > 0$ above $2T_c$. This effect is associated with the evaporation of the magnetic component of the gluon plasma~\cite{Chernodub_2006gu, Liao_2008jg} and the associated string dynamics~\cite{Chernodub_2006gu, Glozman_2022zpy}. Thus, the negative-valued condensate in 
$I_{\mathrm{magn}}$ should nullify the positive contribution of the correlator (mechanical) term $I_{\mathrm{mech}}$ in at a certain temperature $T_s$ below $2 T_c$ in agreement with our estimate~\eq{eq_T_s}.

The suggested mechanism is also in qualitative agreement with previous numerical observations indicating that {\it rigid} rotation increases the critical transition temperature $T_c$~\cite{Braguta_2020biu, Braguta_2021jgn}. Indeed, if the rigid rotation makes the plasma colder, then stronger thermal fluctuations (and, consequently, higher temperatures) are needed to destroy the confinement phase in the rotating plasma as compared to the non-rotating plasma. This simple observation explains the effect of raising critical temperature $T_c$ with increasing angular frequency $\Omega$. Moreover, the crucial role of the magnetic condensate in our mechanism suggests that this effect should be absent for non-gluonic degrees of freedom. The latter hypothesis is perfectly consistent with the recent first-principle observation made separately for quarks and gluons in Ref.~\cite{Braguta_2022str} 

To clarify the contribution of quarks to the moment of inertia, we notice that Eq.~(\ref{eq_I_F}) remains also valid in QCD. Namely, the total angular momentum $M^{12}$ now includes not only the gluon part (\ref{gl_ang_mom}), but also the orbital, $\bar \psi \gamma_4 (x D_y - y  D_x) \psi$, and spin, $i/2 \bar \psi \gamma_4 \sigma_{12} \psi$, angular momenta of quarks. While the quark fields make a positive thermal contribution to the mechanical term $I_{\mathrm{mech}}$, 
the gluomagnetic contribution $I_{\mathrm{magn}}$ stays negative in QCD~\cite{Ishii:1994ku}. 
Thus, we believe that the rigid rotation of quark-gluon plasma is also unstable in a region near $T_c$.

\begin{figure}[t]
\centering
  \includegraphics[width=.99\linewidth]{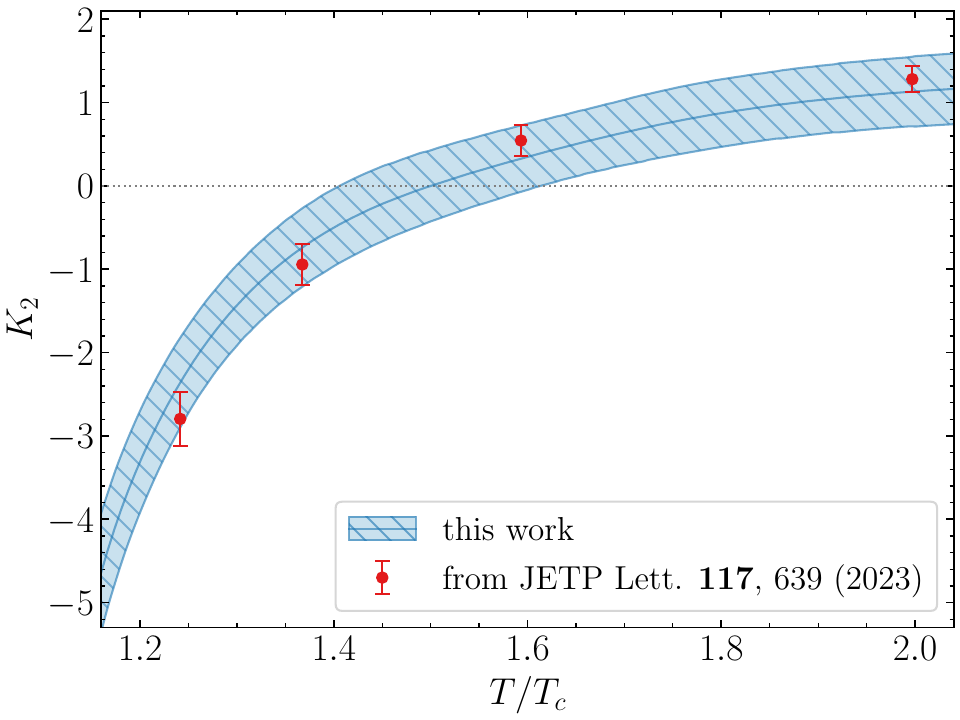}
  \caption{The dimensionless moment of inertia $K_2$ of the gluon plasma as a function of the temperature $T$, obtained in this work after analytic continuation procedure, and the continuum limit results of the direct computation on non-rotating lattices~\cite{Braguta:2023kwl}.}
  \label{fig_K2comp}
\end{figure}

Finally, it is worth mentioning that in classical mechanics, the moment of inertia of a physical body enters various quantities and, consequently, can be calculated using several ways. Likewise, the moment of inertia of the rotating plasma can be determined numerically in independent physical environments. One of these methods, implemented in the present paper, includes the direct calculation of the free energy of a rotating gluon plasma. Another method, used in our earlier numerical calculations on coarser lattices~\cite{Braguta:2023kwl}, applies to a non-rotating gluonic system, in a reminiscence to a classical calculation of the moment of inertia via a volume integral over the mass distribution within the body. The results of these two independent approaches, compared in Fig.~\ref{fig_K2comp}, show an excellent agreement with each other. This coincidence supports the validity of our analytic continuation procedure, and it clarifies our proper understanding of the constant line physics in the evaluation of the free energy of the rotating system~\eq{eq_free_energy}.

%
%

%
%
\section{Conclusions}

All {\it field-theoretical analytical and first\--principle numerical} approaches dedicated to the investigation of the thermodynamics of rotating quark-gluon plasma consider a rigidly rotating plasma, meaning the angular velocity $\Omega$ at all points of the system, regardless of the distance to the rotational axis, takes the same value\footnote{Exceptions in the literature are not known to the authors.}. In other words, the plasma rotates like a solid. 

In our work, we show that below the supervortical temperature~\eq{eq_T_s}, the rigid rotation of the gluon plasma is thermodynamically unstable even at slow rotational velocities. This effect exhibits a striking similarity with spinning black holes~\cite{Whiting1988, Prestidge2000, Reall2001, Monteiro_2009tc, Altamirano2014}. While the curved gravitational background promotes the back-hole rotational instability, the instability in the gluon plasma originates from the scale anomaly via the thermal magnetic gluon condensate. Thus, we conclude that 
rigid rotation cannot be used for thermodynamic reasons for a realistic study of the rotation of the gluon plasma.

Our results also suggest that the puzzling discrepancy between numerical~\cite{Braguta_2020biu,Braguta_2021jgn,Braguta_2022str} and analytical 
\cite{Chen_2015hfc, Jiang_2016wvv, Chernodub_2016kxh, Chernodub_2017ref, Wang_2018sur, Zhang_2020hha, Sadooghi_2021upd,Chen_2020ath, Zhao:2022uxc, Chernodub_2022veq, Fujimoto_2021xix, Golubtsova:2021agl, Chen:2022smf, Golubtsova:2022ldm, Chernodub_2020qah}
predictions for the critical temperature of rotating QCD (gluon) plasma might originate from the scale anomaly, which should be taken into account appropriately. We demonstrate that the magnetic gluon condensate -- which has a nonperturbative component at any temperature -- plays a crucial role in rotating quark-gluon plasma.

\section*{Acknowledgements}
The authors are grateful to Oleg Teryaev for useful discussions and to Andrey Kotov for the help 
at the initial stage of the project. This work has been carried out using computing resources of the Federal collective usage center Complex for Simulation and Data Processing for Mega-science Facilities at NRC ``Kurchatov Institute'', http://ckp.nrcki.ru/ and the Supercomputer ``Govorun'' of Joint Institute for Nuclear Research. This work was supported by the Russian Science Foundation (project no. 23-12-00072).

\appendix
\section{Supplementary material}

Supplementary material related to this article can be found online at URL.

%
%
%
%
%

\bibliographystyle{elsarticle-num}
\bibliography{plasma}

\clearpage
\appendix
\renewcommand\thefigure{S\arabic{figure}} 
\setcounter{figure}{0}

\renewcommand{\thesection}{SM\arabic{section}}
\setcounter{section}{0}

\renewcommand{\thepage}{S\arabic{page}}
\setcounter{page}{1}

\onecolumn

\centerline{\Large{Supplementary Material}}
\vskip 5mm
\centerline{\bf \large{Negative moment of inertia and rotational instability of gluon plasma}}
\vskip 3mm
\centerline{by Victor V. Braguta, Maxim N. Chernodub, Artem A. Roenko, Dmitrii A. Sychev}

\renewcommand\theequation{S1.\arabic{equation}} 
\setcounter{equation}{0}
\section{Scale (trace) anomaly and equation of state}
\label{app_EoS}
To calculate the free energy density $f = F/V$, we use the standard integral method~\cite{Boyd_1996bx}
\begin{subequations}
\label{eqSM_F_S_lat}
\begin{align}
    \frac{f(T)}{T^4} & = - N_t^4 \int_{\beta_0}^{\beta} d \beta' \Delta s(\beta')\,,
    \label{eqSM_F_lattice} \\
    \Delta s(\beta) & = \langle s(\beta) \rangle_{T=0} - \langle s(\beta) \rangle_{T},
    \label{eqSM_Delta_S}
\end{align} 
\end{subequations}
where the $\mathrm{SU}(N_c)$ lattice coupling $\beta = 2 N_c/g_0^2$ is expressed in terms of the bare continuum coupling $g_0$, while $s=T/V\cdot S$ is the density of the lattice action $S$. This approach directly follows from the common thermodynamic relation $F/T = - \ln Z$ and its derivative with respect to the lattice coupling $\beta$. 
The free energy density also relates to the scale (trace) anomaly~\cite{Boyd_1996bx}:
\begin{align}
    \langle T^{\mu}_{\ \mu}\rangle  = - T^5 \frac{d}{dT}\left(\frac{f}{T^4}\right) \,.
    \label{eqSM_theta_P}
\end{align}
A justification of the above formula for a rotating system will be given in Section~\ref{app_trace_rotating}.

Integrating~\eq{eqSM_theta_P}, we get the free energy density:
\begin{equation}
    \frac{f(T)}{T^4} = -\int_{0}^T \frac{d T'}{T'} \frac{\langle T^{\mu}_{\ \mu}\rangle(T')}{{T'}^4}\,, 
    \label{eqSM_F_int}
\end{equation}
highlighting the {importance} of the anomaly $\langle T^{\mu}_{\ \mu}\rangle \neq 0$. When defining the integral in Eq.~\eq{eqSM_F_int}, we used the fact that the anomalous trace vanishes rapidly at low temperatures, $\langle T^{\mu}_{\ \mu}\rangle \sim \exp( - M/T)$, due to the mass gap $M \neq 0$. The lattice formula~\eq{eqSM_F_lattice} has the same meaning as the continuum relation~\eq{eqSM_F_int}, with the right-hand-side of Eq.~\eq{eqSM_F_lattice} expressed via the lattice form of the scale anomaly~\cite{Boyd_1996bx}:
\begin{equation}\label{eqSM_Tmm_vs_ds}
    \frac{\langle T^{\mu}_{\,\mu}\rangle}{T^4} = - N_t^4 a \frac{d\beta}{da} \Delta s,
\end{equation}
where the $\beta$-function $a\cdot d\beta/da = - T\cdot d\beta/dT$ is computable via the scale dependence of the lattice spacing $a = a(\beta)$.

The lattice formula~\eq{eqSM_F_lattice} is also suitable for direct calculation of the free energy density $f$ in the non-inertial co-rotating reference frame. The method, described by Eqs.~\eqref{eqSM_F_S_lat}, remains valid for the rotating lattices if the angular velocity in lattice units is kept constant with the variation in $\beta$, which corresponds to the constant linear velocity at the boundary of the rotating system.
In Eq.~\eq{eqSM_F_lattice}, the lower integration limit $\beta_0$ is chosen in a deep confinement phase where the integrand, represented by the difference~\eq{eqSM_Delta_S} in the expectation values of the action at vanishing and finite temperatures, is negligibly small.

As explained in the main text, the response of the co-rotating free energy to a slow rotation may be expressed in terms of the moment of inertia:
\begin{equation}
    F(T, R, \Omega) = F_0(T, R) - \frac{1}{2} I(T,R) \Omega^2 \,, \label{eqSM_F_series}
\end{equation}
where $\Omega$ is the angular velocity. As one can see from Eqs.~\eqref{eqSM_F_int}, and~\eqref{eqSM_F_series}: 
\begin{equation}
    I(T) = - V T^4 \int_{0}^T \frac{d T'}{T'} \frac{\langle T^{\mu}_{\ \mu}\rangle^{(2)}(T')}{{T'}^4}\,, 
    \label{eqSM_I_int}
\end{equation}
where $\langle T^{\mu}_{\ \mu}\rangle^{(2)}$ is the second moment of the anomalous trace
\begin{equation}
    \label{eqSM_T_moments}
    \langle T^{\mu}_{\ \mu}\rangle^{(2)}(T) = - \left[ \frac{\partial^{2}}{\partial \Omega^{2}} \langle T^{\mu}_{\ \mu}\rangle(T,\Omega) \right]{\Biggl|}_{\Omega = 0}\,.
\end{equation}
The equation~\eq{eqSM_T_moments} can be rewritten in terms of the imaginary angular velocity using the correspondence $\Omega_I = - i \Omega$ (i.e. $\partial^2/\partial\Omega^2 = - \partial^2/\partial\Omega_I^2$). Therefore, the moment of inertia is connected to the rotational response of the scale anomaly. 

\renewcommand\theequation{S2.\arabic{equation}} 
\setcounter{equation}{0}
\section{Scale (trace) anomaly for a rotating system}
\label{app_trace_rotating}

Our derivation of the equation of state and associated nontrivial inertial properties of gluon plasma is based on the specific equation for the trace (conformal) anomaly~\eqref{eqSM_theta_P}. This relation is obviously valid in the static (non-rotating) plasma, while its applicability to the rotating system is not evident. Below, we show that Eq.~\eqref{eqSM_theta_P} is also applicable to the plasma in a rotating cylinder. 

\subsection{$\Omega R$-scaling of free energy}
\label{app_Omega_R}

The first step of our arguments takes into account the fact that the system in rigid rotation must be spatially bounded in the transversal plane to respect causality. A particular geometry of the transversal cross-section does not play a role as one can demonstrate that a difference in shapes (e.g., round vs. square) results only in an overall geometrical factor that does not alter the functional form of Eq.~\eqref{eqSM_theta_P}. For simplicity, we consider below the round shape of the cylinder with the radius $R$. We assume that the height of the cylinder is infinite as it does not affect our discussion on rotation, for which only the physics in the transverse plane plays a role.

The second observation is that Eqs.~\eqref{eq_I_T_Omega} and \eqref{eq_F_all} imply that the free energy, in the quadratic order, depends on the angular frequency $\Omega$ only through the velocity of plasma at the boundary $v_R = \Omega R$, Eq.~\eqref{eq_V_I_R}. In other words, for the free energy density, the following relation holds:
\begin{align}
    f(T,\Omega,R) = f(T,\Omega R)\,.
\label{eq_f_Omega_R}
\end{align}
For shortness, we call this property the $v_R$-scaling below. Notice that the statement of the $v_R$-scaling is intuitively nontrivial despite its simplicity. 

To justify the relevance of Eq.~\eq{eq_f_Omega_R} to our discussion, we mention that the moment of inertia is a property revealed in the quadratic order of $\Omega$. We remind that Eqs.~\eqref{eq_I_T_Omega} and \eqref{eq_F_all} appear on the basis of generic analytical arguments implying that the first correction to the energy density~\eq{eq_f_Omega_R} appears in the quadratic order in $\Omega$. 

We also mention, in bypassing, that the quadratic order is valid for moderately non-relativistic velocities corresponding to rotating quark-gluon plasma observed at RHIC (for which $v_R^2 \ll 1$). While the statement of the $v_R$-scaling in quadratic order of $\Omega$ is sufficient for the purpose of this paper, we believe that it is valid in all orders of $\Omega$, which adds more generality to our approach. 

\subsection{Scale (trace) anomaly in a rotating cylinder}
\label{app_trace}

Technically, the expression for the moment of inertia~\eqref{eq_I_int} originates from the integral formula for free energy  relation~\eqref{eqSM_F_int}, which, in turn, appears from the equation for the trace anomaly in the non-rotating state in thermodynamic limit:
\begin{align}
\left(4 - T \frac{\partial}{\partial T}\right) f(T) = \bigl\langle T^\mu_{\,\mu}\bigr\rangle \,, \qquad \text{[thermodynamic limit, no rotation]}\,.
\label{eq_trace_thermodynamic}
\end{align}
The coefficient $4$ indicates the canonical dimensions of the free energy density, $[f] = [\text{mass}]^4$. Equation~\eqref{eq_trace_thermodynamic} can also be rewritten in the concise form of Eq.~\eqref{eqSM_theta_P}, from which Eq.~\eqref{eqSM_F_int} follows immediately~\cite{Boyd_1996bx}. 

In the presence of other thermodynamic variables characterizing the system (such as a chemical potential $\mu$, size $l$, etc.), the logarithmic temperature gradient in Eq.~\eq{eq_trace_thermodynamic} should be extended as 
\begin{align}
    T \frac{\partial}{\partial T} \to T \frac{\partial}{\partial T} + \sum_{a} d_a X_a \frac{\partial}{\partial X_a}\,,
    \label{eq_T_to}
\end{align}
where $d_a$ is the dimension of the quantity $X_a$ in the dimension of a mass scale ($d_\mu = + 1$ for $\mu$ and $d_l = - 1$ for $l$, etc).

In a cylinder of the radius $R$ (with $d_R = -1$) rotating with the angular frequency $\Omega$ (with $d_\Omega = + 1$), the extension~\eqref{eq_T_to} applied to Eq.~\eq{eq_trace_thermodynamic} gives us the corresponding differential anomaly equation:
\begin{align}
\left(4 - T \frac{\partial}{\partial T} - \Omega \frac{\partial}{\partial \Omega}  + R \frac{\partial}{\partial R}  \right) f(T) = \bigl\langle T^\mu_{\,\mu}\bigr\rangle \,, \qquad \text{[rotating cylinder]}\,.
\label{eq_trace_cylinder}
\end{align}
However, the $v_R$-scaling~\eqref{eq_f_Omega_R}, established earlier, implies that the effects of the rigid rotation and the finite transverse size of the rotating cylinder cancel each other in the trance anomaly equation~\eq{eq_trace_cylinder}, which brings us back to Eq.~\eq{eq_trace_thermodynamic}, thus proving, consequently, Eq.~\eqref{eqSM_F_int} and our main formula~\eq{eqSM_I_int}.

\renewcommand\theequation{S3.\arabic{equation}} 
\setcounter{equation}{0}
\section{Lattice action and simulation details}
\label{app_simulations}
To perform the simulations, the action $S$ of rotating gluon fields should be formulated in the curved background of the co-rotating frame~\cite{Yamamoto_2013zwa}:
\begin{equation}
g^E_{\mu \nu} = 
\begin{pmatrix}
1 & 0 & 0 & x_2 \Omega_I \\
0 & 1 & 0 & -x_1 \Omega_I  \\ 
0 & 0 & 1 & 0 \\
x_2\Omega_I  & - x_1 \Omega_I & 0 & 1 + x_\perp^2 \Omega_I^2
\end{pmatrix}\,,
\label{eqSM_g_E}
\end{equation}
written in the Euclidean coordinates $x^\mu = (x^1,\dots,x^4)$, where $x_4 = - i t$ is the imaginary time coordinate, and $x^2_\perp = x^2_1 + x_2^2$. The system rotates around the $x_3$ axis. 

The angular velocity in Eq.~\eq{eqSM_g_E} is put in the purely imaginary form $\Omega_I = - i \Omega$ to avoid the sign problem~\cite{Yamamoto_2013zwa}. In particular, the velocity $v_R=\Omega R$ at the boundary, where $R$ is the distance to the rotational axis, 
becomes imaginary $v_I = - i v_R$. 
The expressions for the Minkowski spacetime can be obtained by the analytic continuation. 

The gluon action in the co-rotating frame in the continuum Euclidean spacetime has the following form:
\begin{align}
    S = \frac{1}{4 g_0^2} \int\! d^4 x\, \sqrt{g_E}\,  g_E^{\mu \nu} g_E^{\alpha \beta} F_{\mu \alpha}^{a} F_{\nu \beta}^{a} \,,
    \label{eqSM_action_curved}
\end{align}
where $g^E_{\mu \nu} = (g_E^{\mu \nu})^{-1}$ is the rotational Euclidean metric~\eq{eqSM_g_E} with the determinant $g_E = {\mathrm{det}}\, (g_{\mu\nu})= 1$ and $F^a_{\mu\nu}$ is the field strength of SU(3) gauge field. 

We discretize rotating terms in the action~\eq{eqSM_action_curved} following Ref.~\cite{Yamamoto_2013zwa, Braguta_2021jgn} and use the tree-level improved Symanzik gauge action for the  terms without rotation~\cite{Curci:1983an, Luscher:1985zq}: 
\begin{multline}\label{eqSM:rot_action_lat_imp}
S_{G} = \beta  \sum_{x}\Big( (c_0 + x_\perp^{2}\Omega_I^{2}) (1 - \frac{1}{N_c} \text{Re} \Tr \ \bar{U}_{12} ) + (c_0 + x_2^{2}\Omega_I^{2}) (1 - \frac{1}{N_c} \text{Re} \Tr \ \bar{U}_{13} )+{} \\
{}+ (c_0 + x_1^{2}\Omega_I^{2}) (1 - \frac{1}{N_c} \text{Re} \Tr \ \bar{U}_{23} ) +
c_0 \big(3 - \frac{1}{N_c} \text{Re} \Tr \ 
( \bar{U}_{14} + \bar{U}_{24} + \bar{U}_{34})\big) -{} \\
{} + \sum_{\mu\neq\nu} c_1 (1 - \frac{1}{N_c} \text{Re} \Tr \ \bar{W}_{\mu\nu}^{1\times2})- \frac{1}{N_c} \text{Re} 
\Tr \ \big( x_2\Omega_I (\bar{V}_{124} + \bar{V}_{134})
- x_1\Omega_I ( \bar{V}_{214} +  \bar{V}_{234}) 
+ x_1 x_2\Omega_I^{2}  \bar{V}_{132}\big) \Big)\, ,
\end{multline}
where $\bar{U}_{\mu\nu}$ denotes the clover-type average of four plaquettes, $\bar{W}_{\mu\nu}^{1\times2}$ is the rectangular loop,
$\bar{V}_{\mu\nu\rho}$ is the asymmetric chair-type average of eight chairs~\cite{Braguta_2021jgn}, and $c_0 = 1 - 8 c_1$, $c_1 = -1/12$. The action \eqref{eqSM:rot_action_lat_imp} in the case $c_1 = 0$ coincides with the lattice gauge action used in Refs.~\cite{Braguta_2020biu, Braguta_2021jgn}.

For each lattice size, we keep the (imaginary) angular velocity in lattice units unchanged with the variation of $\beta$. Therefore, the linear velocity $v_I$ at the boundary of the system remains constant with the changes in temperature. In our simulations the linear velocity takes the following values: $v_I^2 = 0.000, 0.015, 0.030, 0.045, 0.060, 0.075, 0.090$.

As mentioned in the main text, the absence of massless excitations in the deconfinement phase implies the independence of our results on the type of boundary conditions in the transverse spatial directions, provided the spatial volume is large enough. To verify this property, we additionally calculate the normalized moment of inertia $K_2 = -I / (F_0 R^2)$ for the system with open boundary conditions and find an agreement with the periodic lattices, albeit more significant uncertainties. The corresponding supervortical temperature for the open system, $T_s/T_c = 1.53(15)$, agrees with the estimate for periodic boundary conditions, $T_s/T_c = 1.50(10)$.

In order to estimate the systematic errors related to our determination of the supervortical temperature~$T_s$, we use 
several methods of numerical integration in Eq.~\eqref{eqSM_F_lattice} and several upper limits $v_I^{\text{(max)}}$ for the fit of the free energy density with a quadratic function in $v_I$. 
The estimated uncertainty in the final result for the supervortical temperature incorporates both statistical and systematic contributions.

To set the temperature scale, we use the results for the string tension from Ref.~\cite{Beinlich:1997ia}.
The definition of temperature in the background gravitational field is not trivial. The Tolman-Ehrenfest law implies that local temperature $T(r)$  in a system in thermodynamic equilibrium subjected to a static inhomogeneous gravitational field depends on the spatial coordinate according to the relation $T(r) \sqrt{g_{tt}} = T = \textit{const}$~\cite{Tolman_1930ona, Tolman_1930zza, Braguta_2021jgn}, which is written in the mostly-minus metric convention. Following Ref.~\cite{Braguta_2021jgn}, we avoid ambiguities in the definition of temperature by always using its value at the rotation axis $T(0) \equiv T$, where the metric-related effects are absent. The on-axis temperature is directly related to the length of the compactified imaginary-time direction of the Euclidean time $1/T = a(\beta) N_t$  in Eq.~\eqref{eqSM_action_curved} and exactly coincides with the temperature value for the non-rotating system with a flat metric at the same lattice parameters.
For the non-rotating lattices with periodic boundary conditions, we use the values of the critical coupling $\beta_c$ taken from Ref.~\cite{Beinlich:1997ia}. For the case of open boundary conditions, we determine $\beta_c$ from the peak of the Polyakov loop susceptibility.

Simulations are performed using Monte Carlo algorithm, each sweep consists of one heatbath update and two steps of the overrelaxation update. In finite (zero) temperature simulations, typical statistics of about 5000-40000 (2000-10000) sweeps after thermalization for each set of parameters are employed. The statistical uncertainties are estimated via the jackknife method.

\end{document}